\documentclass[aps,prb,reprint,superscriptaddress]{revtex4-1}
\usepackage{amsmath}
\usepackage{graphicx}% Include figure files
\usepackage{color}

\renewcommand{\Im}{\operatorname{Im}}

\newcommand{\figref}[1]{Fig.~\ref{#1}}
\newcommand{\Figref}[1]{Figure~\ref{#1}}

\renewcommand{\eqref}[1]{Eq.~(\ref{#1})}

\begin{document}
\title{Near-field thermal radiation transfer controlled by plasmons in graphene}

\author{Ognjen Ilic}
\email{ilico@mit.edu}
\affiliation{
  Department of Physics, Massachusetts Institute of Technology,
  77 Massachusetts Avenue, Cambridge, Massachusetts, 02139, USA
}
\author{Marinko Jablan}
\affiliation{
  Department of Physics, University of Zagreb, 
  Bijeni\v{c}ka c. 32, 10000 Zagreb, Croatia
}
\author {John D. Joannopoulos}
\affiliation{
  Department of Physics, Massachusetts Institute of Technology,
  77 Massachusetts Avenue, Cambridge, Massachusetts, 02139, USA
}
\author{Ivan Celanovic}
\affiliation{
  Institute for Soldier Nanotechnologies, Massachusetts Institute of Technology,
  77 Massachusetts Avenue, Cambridge, Massachusetts 02139, USA \\
}
\author{Hrvoje Buljan}
\affiliation{
  Department of Physics, University of Zagreb, 
  Bijeni\v{c}ka c. 32, 10000 Zagreb, Croatia
}
\author {Marin Solja\v{c}i\'{c}}
\affiliation{
  Department of Physics, Massachusetts Institute of Technology,
  77 Massachusetts Avenue, Cambridge, Massachusetts, 02139, USA
}


\begin{abstract}
  It is shown that thermally excited plasmon-polariton modes can
  strongly mediate, enhance and \emph{tune} the near-field radiation
  transfer between two closely separated graphene sheets. The dependence
  of near-field heat exchange on doping and electron relaxation
  time is analyzed in the near infra-red within the framework of
  fluctuational electrodynamics. The dominant contribution to heat
  transfer can be controlled to arise from either interband or intraband
  processes. We predict maximum transfer at low doping and for plasmons
  in two graphene sheets in resonance, with orders-of-magnitude
  enhancement (e.g. $10^2$ to $10^3$ for separations between $0.1\mu m$
  to $10nm$) over the Stefan-Boltzmann law, known as the far field
  limit. Strong, tunable, near-field transfer offers the promise of an
  externally controllable thermal switch as well as a novel hybrid
  graphene-graphene thermoelectric/thermophotovoltaic energy conversion
  platform.
\end{abstract}

                             % 44.40.+a : Thermal Radiation
                             % 78.67.Wj : Optical properties of Graphene
                             % 73.20.Mf : Collective excitations (plasmons)
			
\maketitle

\noindent Heat transfer between two bodies can be greatly enhanced in
the \emph{near field}, i.e. by bringing their surfaces close together
to allow tunnelling of evanescent photon modes. For two parallel,
semi-infinite, dielectric surfaces of index of refraction $n$, maximum
flux enhancement is known to be $n^2$ times the Planck's black body
limit \cite{Landau1980}. However, particularly interesting near-field
radiation transfer phenomena involve thermal excitation of various
surface modes. Due to their localization and evanescent nature, it
is only at sub-micron separations that these modes become relevant.
Measuring near-field transfer has been experimentally difficult
\cite{Hargreaves1969,Narayanaswamy2008a,Shen2009a,Rousseau2009};
nevertheless, the promise of order-of-magnitude enhancement over the
far field Planck's black body limit has made near-field transfer
the topic of much research \cite{Joulain2005}. A promising
class of materials for enhancing the near-field transfer are plasmonic
materials, due to high density of modes around the frequency of
plasmons. The potential of graphene \cite{Novoselov2004} as a
versatile and tunable plasmonic material has already been recognized
in applications such as teraherz optoelectronics and transformation
optics \cite{Jablan2009,Wunsch2006,Hwang2007,Ju2011,Vakil2011}. Unlike
in metals, where high plasma frequencies make thermal excitation of
surface modes difficult, plasmon frequencies in graphene can be anywhere
from the teraherz to the near infra-red \cite{Rana2011}. In addition,
the dependence of graphene conductivity on chemical potential, which
in turn can be controlled by doping or by gating, allows for a tunable
plasmonic dispersion relation. Transfer between graphene and amorphous
$\textrm{SiO}_2$ \cite{Persson2010,Volokitin2011} as well as application
of graphene as a thermal emitter in a near-field thermophotovoltaic
(TPV) system has been reported \cite{Ilic2012}. Here we analyze the
contribution of plasmon-polaritons to graphene-graphene near-field heat
transfer. The choice of identical coupled systems is predicated
on the idea that resonant enhancement could lead to even greater heat
transfer capacity. Indeed, we find maximal transfer for resonantly
coupled plasmon modes (corresponding to similar doping in the two
graphene sheets), which can be orders of magnitude larger than the heat
transfer between two black bodies in the far field.

In general, the radiative heat transfer between two bodies at temperatures
$T_1$ and $T_2$ is given by
\begin{equation}
  H= 
  \int_0^{\infty} d\omega \left[\Theta(\omega,T_1)-\Theta(\omega,T_2)\right]
  f(\omega;T_1,T_2) 
  \label{eq:spectral} 
\end{equation}
where $\Theta(\omega,T)=\hbar\omega/(e^{\hbar\omega/k_bT}-1)$ is the
average energy of a photon at frequency $\omega$ (the Boltzmann factor),
and $f(\omega;T_1,T_2)$ is the \emph{spectral transfer function},
characterizing frequency dependence of the heat exchange (i.e. how
much heat is exchanged at a given frequency). In the context of
fluctuational electrodynamics \cite{Rytov1987}, the spectral transfer
function $f(\omega;T_1,T_2)$ is calculated in the following way:
thermal fluctuations in the first (emitter) medium induce correlations
between electric currents, which are proportional to the real part of
the medium conductivity \cite{Haus1961}; next, using Green functions,
we can find the electromagnetic fields in the second (absorber) medium
induced by the fluctuating currents in the first \cite{Sipe1987};
finally, the radiation transfer is obtained by calculating the
Poynting flux around (or the ohmic losses within) the second medium.
This approach has been used to numerically calculate the near-field
transfer between two half-spaces \cite{Rytov1987,Pendry1999a}, as
well as generalizations such as two slabs \cite{Ben-Abdallah2009},
sphere and a plane \cite{Mulet2002,Narayanaswamy2008a} two spheres
\cite{Narayanaswamy2008}, as well as 1D periodic structures
\cite{Rodriguez2011}.

\begin{figure}[htb] 
\centerline{\includegraphics{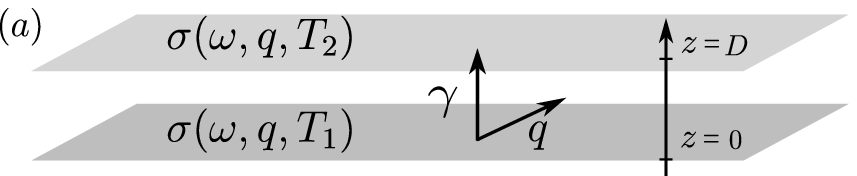}}
\centerline{\includegraphics{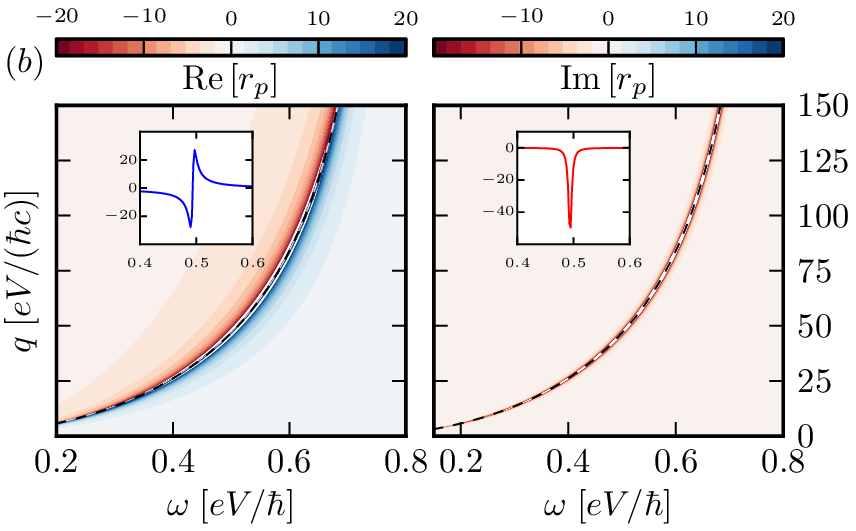}}
\caption{
  (a) Schematic diagram of the radiation transfer problem: a suspended
  sheet of graphene at temperature $T_1$ is radiating to another
  suspended graphene sheet at temperature $T_2$ and distance $D$
  away. k-vector components are $q$,$\gamma$, for the parallel and
  perpendicular component, respectively. (b) Real and imaginary parts
  of graphene $p$-polarization reflection coefficient for $\mu=0.5eV$,
  $T=300K$, $\tau=10^{-13}s$. Dashed line is the vacuum plasmon
  dispersion relation (\ref{eq:disp-vac}) for the graphene sheet.
  Insets show the real and imaginary part of reflectivity at $q\approx
  30eV/\hbar c$ as a function of $\omega$.}
\label{fig:sketch}
\end{figure}

The system we analyze, shown in \figref{fig:sketch}, consists of a
suspended graphene sheet at temperature $T_1$ emitting to another
suspended graphene sheet held at room temperature $T_2=300K$, and a
distance $D$ away. In general, the $p$-polarization spectral transfer
function for evanescent modes between two bodies is
\begin{align}
  f_p(\omega;T_1,T_2) = \frac{1}{\pi^2} \int_{\omega/c}^{\infty} dq q
  \frac{\Im(r_1^p)\Im(r_2^p)}{\left| 1- r_1^pr_2^pe^{2i\gamma D}\right|^2}
  e^{2i\gamma D}
\label{eq:spectral-gg}
\end{align}
where $\gamma=\sqrt{\omega^2/c^2 - q^2}$ is the perpendicular
wave-vector and $r_{1(2)}$ is the reflection coefficient for the
bottom(top) body; note that $r_{1,2}$ depend on $T$, and hence the
$T$-dependence of $f(\omega,T_1,T_2)$. Integration is over the parallel
wave-vector $q$, limited only to the evanescent ($q> \omega/c$) modes.
The spectral transfer function (\ref{eq:spectral-gg}) was derived for
the case of two semi-infinite slabs \cite{Joulain2005}; however, it can
be shown that the same expression is valid when any of the two bodies is
a 2D system, such as graphene \cite{Ilic2012}.
Since graphene absorbs poorly ($2.3\%$) in the far
field (hence is also a poor emitter), not including the propagating
modes is a good approximation. The contribution of evanescent $s$-polarized
modes can also be calculated using \eqref{eq:spectral-gg}, but it turns
out to be negligible compared to $p$-polarized modes, as we discuss
later. We assume graphene is completely characterized by its complex
optical conductivity $\sigma=\sigma_r + i\sigma_i$, which depends
on angular frequency $\omega$, electron scattering lifetime $\tau$,
chemical potential $\mu$, and temperature $T$. Furthermore, the graphene
conductivity is taken to be independent of the parallel wave-vector
$q$ (see discussion below), and consists of the Drude (intraband) and
interband conductivity, expressed respectively as \cite{Falkovsky2008}
\begin{align} 
  \sigma_{D} &= \frac{i}{\omega+i/\tau}\frac{e^2 2k_{b}T}{\pi\hbar^2}
    \textrm{ln}\left[2\textrm{cosh}\frac{\mu}{2k_bT}\right]
  \label{eq:cond}
  \\
  \sigma_{I} &= \frac{e^2}{4\hbar}
    \left[G\left(\frac{\hbar\omega}{2}\right) + i
    \frac{4\hbar\omega}{\pi}\int_0^{\infty}
    \frac{G(\xi)-G(\hbar\omega/2)}
      {(\hbar\omega)^2-4\xi^2}d\xi
    \right] \nonumber
\end{align} 
where $G(\xi) =\textrm{sinh}(\xi/k_bT)/(\textrm{cosh}
(\mu/k_bT)+\textrm{cosh}(\xi/k_bT))$, and $\mu$ is the chemical
potential. Various electron scattering processes are taken into account
through the relaxation time $\tau$. From DC mobility measurements in
graphene, one obtains \cite{Jablan2009} an order-of-magnitude value of
$\tau\approx 10^{-13}s$.

First we discuss the electrodynamic properties of a single suspended
sheet of graphene, inherent in the $p$-polarization reflection
coefficient, which is illustrated in \Figref{fig:sketch}b. The
reflection coefficient is $r_p=(1-\epsilon)/\epsilon$, where
$\epsilon=1+\gamma\sigma/(2\epsilon_0\omega)$ is the dielectric function
of graphene \cite{Falkovsky2008}. Its pole $\epsilon=0$ corresponds to
the dispersion relation of $p$-polarized plasmon modes \cite{Jablan2009}
\begin{equation}
  q = \epsilon_0 \frac{2i\omega}{\sigma(\omega,T)},
\label{eq:disp-vac}
\end{equation}
which is shown as the dashed line in \figref{fig:sketch}b.
\Figref{fig:sketch} shows plasmons exist in a strongly non-retarded
regime ($q\gg\omega/c$), indicating a tightly confined plasmon polariton
mode. Graphene also supports $s$-polarized surface modes with a
dispersion relation very close to the light line \cite{Mikhailov2007}.
However, due to the large density of states and the tightly confined
nature of $p$-polarized surface modes, it is the $p$-polarization that
dominates (as our calculations confirm) the near-field transfer.

\begin{figure}[htb] 
\centerline{\includegraphics{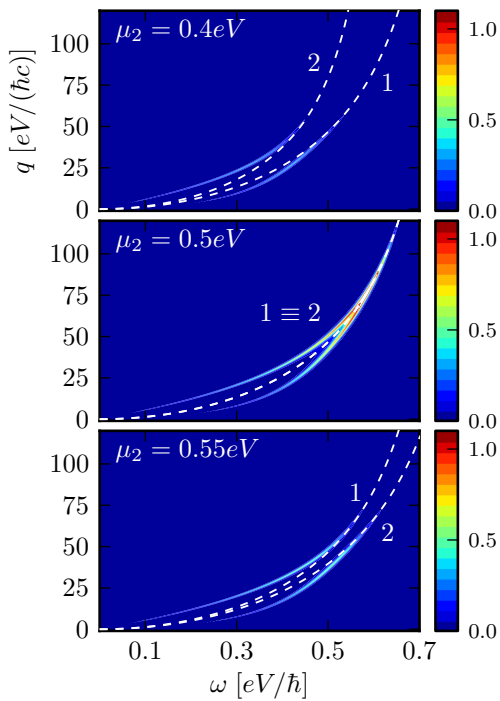}}
\caption{
  Contour plot of the integrand (a.u.) in $f_p(\omega)$ from
  (\ref{eq:spectral-gg}), for two graphene sheets at $T_{1,2}=300K$,
  separated by $D=10nm$. Chemical potentials are $\mu_1=0.5eV$, while
  $\mu_2$ is different for each plot. Dashed lines correspond to the
  vacuum plasmon dispersion relations for the bottom (1) and the
  top (2) graphene sheet.}
\label{fig:2}
\end{figure}

When two parallel graphene sheets are sufficiently close (see
\figref{fig:sketch}a), their plasmonic modes can become coupled. The
dispersion of these coupled modes is $1-r_1^p(\omega)r_2^p(\omega)
e^{-2 q D}=0$, when $q\gg\omega/c$, so $\gamma\approx iq$, which is
exactly the pole of the integrand of the spectral transfer function
(\ref{eq:spectral-gg}). The integrand is illustrated in \Figref{fig:2}
for different values of chemical potential. The coupling of modes
is strongest when both graphene sheets have identical parameters
(middle panel in \figref{fig:2}). In that case, their individual
dispersions are identical. Nevertheless, the dispersion of the combined
system shows two branches that dominate the near-field spectral
transfer, i.e. the implicit equation $1-r(\omega)^2e^{-2qD}=0$ for
$\omega(q)$ has two explicit solutions: $\omega_{\mathrm{even}}(q)$ and
$\omega_{\mathrm{odd}}(q)$ for the even, and the odd mode, respectively.
The splitting of two superimposed resonances is
particularly noticeable at smaller wave vectors $q$. For larger $q$,
the splitting disappears, and the resonant matching of peaks of
$\Im(r_{1,2})$ significantly enhances the near-field transfer. As the
chemical potential of one of the sheets changes (top and bottom panel
in \figref{fig:2}), the plasmons in the two sheets move out of resonance,
coupling decreases, the peaks in the integrand approach the individual
(vacuum) plasmons dispersion curves, and the heat transfer becomes lower
than in resonance.

\begin{figure}[htb] 
\centerline{\includegraphics{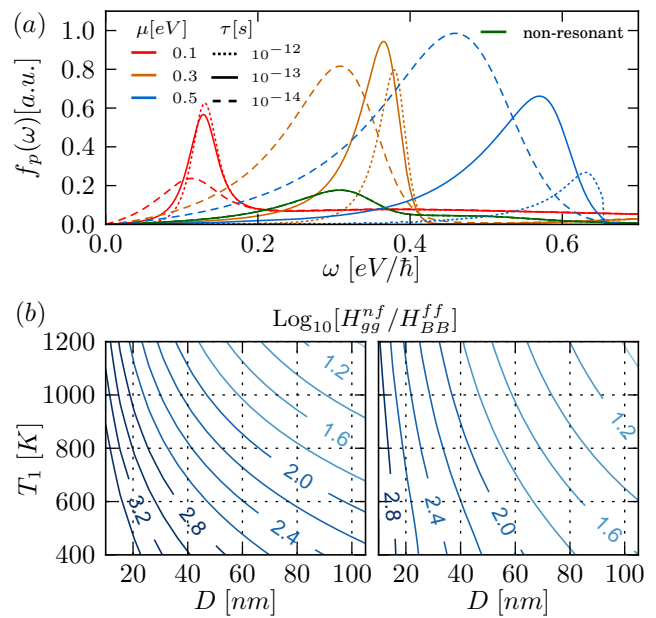}}
\caption{
  (a) Spectral transfer function $f_p(\omega)$ from
  (\ref{eq:spectral-gg}), for plasmons in two graphene sheets at
  resonance, $\mu_{1,2}=\mu$, $\tau_{1,2}=\tau$; $T_{1,2}=300K$,
  $D=10nm$. Solid green line corresponds to the $\mu_{1(2)}=0.3(0.5)eV$,
  $\tau_{1(2)}=10^{-13}(10^{-14})s$ case.
  (b) Contour plot of the integrated ratio of the near-field transfer
  between two graphene sheets, $H_{gg}^{nf}$, and the far field
  transfer between two black bodies, $H_{BB}^{ff}$ for plasmons in resonance
  (left, $\mu_{1,2}=0.1eV$) and out of resonance (right,
  $\mu_{1(2)}=0.1(0.3)eV$). Here, $T_2=300K,\tau_{1,2}=10^{-13}s$.}
\label{fig:3}
\end{figure}

\Figref{fig:3}a shows a highly tunable spectral transfer function $f_p$
for different values of chemical potential and relaxation time.
Given the chemical potential, the relaxation time determines which
processes (interband or intraband) are responsible for the peaks in
spectral transfer. Since interband processes are dominant at high
frequencies, all $\tau$ curves converge in the high frequency limit,
where Drude losses are negligible. However, interband processes can play
a leading role even below the absorption threshold $\omega\approx 2\mu$,
particularly for small chemical potential where thermal broadening
of the interband threshold (on the order of few $k_bT$) becomes more
significant. For example, for $\mu_{1,2}=0.1eV$ (first peak
in \figref{fig:3}a) the similarity between $\tau=10^{-12}s$
and $\tau=10^{-13}s$ spectral transfer functions indicates that the
majority of loss in graphene comes from interband processes. On the
other hand, the Drude (intraband) loss term, usually important for
$\omega<\mu$, can become dominant at higher frequencies, for large
enough $\mu$ (third peak). Finally, a combination of two loss processes,
$\mu_{1(2)}=0.3(0.5)eV$, $\tau_{1(2)}=10^{-13}(10^{-14})s$ can lead to
a hybrid spectral transfer. While the use of $q$-independent
expression for graphene conductivity \eqref{eq:cond} for intraband
processes is a good approximation \cite{Jablan2009}, one must take care
when applying \eqref{eq:cond} to interband transitions. As indicated in
\figref{fig:3}a, interband transitions can play a significant role in
near-field transfer at low doping levels. Here, the contribution from
the non-zero wave-vector becomes important since it broadens the
interband threshold from $2\mu$ to $\sim2\mu-\hbar qv_F$. On the other
hand, this is similar to non-zero temperature effects which also
broaden the interband threshold, so we do not expect a qualitatively
different result with $q$-dependent conductivity.

We quantify the heat exchange in the near-field by plotting
(\figref{fig:3}b) the integrated transfer $H$ from \eqref{eq:spectral}
normalized to the transfer between two black bodies in the far field.
Factoring in the temperature dependence shifts the majority of the
near-field transfer to lower frequencies, due to the exponentially
decaying Boltzmann factor. This implies that while doping or gating
might be advantageous in some applications (for example, emitter-PV cell
bandgap frequency matching in near-field TPV systems \cite{Ilic2012}),
near-field transfer between two graphene sheets is maximized for small
values of doping, despite the stronger peak in spectral transfer
for $\mu_{1,2}=0.3eV$ vs. $\mu_{1,2}=0.1eV$ (\figref{fig:3}a).
For plasmons in resonance with $\mu_{1,2}=0.1eV$ (left panel,
\figref{fig:3}b), we observe orders-of-magnitude increase in heat
exchange, particularly at small separations ($\times1000$ for
$D=20nm,T_1=800K$), but also at separations as large as $0.1 \mu m$.
At larger separations, we observe (not shown) the shift of the peak
of the spectral transfer function $f_p$ to $\mu_{1,2}=0.1eV$ case
(red line in \figref{fig:3}a), indicating that the coupling between
highly localized, large $q$, modes becomes weaker, and the transfer is
dominated by lower-frequency, less evanescent modes. The heat transfer
depends in a complex fashion on the parameters of the system, and
does not seem to yield a simple functional dependence on the emitter
and absorber temperatures (as is the case for two black bodies).
Nevertheless, there is a relative advantage (\figref{fig:3}b) to
operating at lower temperatures, as the temperature dependence of the
near-field transfer appears to grow slower than the $T^4$ black body
dependence. Finally, we note that the temperature dependence of
conductivity reduces the resonant effect when two graphene sheets are
at different temperatures. This reduction is more pronounced for large
temperature difference, shifting the peak of the spectral transfer on
the order of $k_bT$; however, the relative reduction of the integrated
spectral transfer function is small, with the main temperature dependence
coming from the Boltzmann factor. This efficient heat exchange between
two graphene sheets in the near-field, together with recently reported
advances in hot carrier extraction from graphene \cite{Gabor2011}, may
offer a potential for a novel, hybrid thermophotovoltaic/thermoelectric
solid-state heat-to-electricity conversion platform. In addition,
this material system could pave the way toward an externally
controllable thermal switch behavior, where one can, by means of doping
or gating, tune the resonant coupling between the hot and the cold
side.

The authors would like to acknowledge helpful discussions with Pablo
Jarillo-Herrero, Nathan Gabor, Gang Chen, Alejandro Rodriguez and Steven
Johnson. OI and MS were partially supported by the MIT S3TEC Energy
Research Frontier Center of the Department of Energy under Grant No.
DE-SC0001299. MJ was supported in part by the Croatian Ministry of
Science under Grant No. 119-0000000-1015. This work was also partially
supported by the Army Research Office through the Institute for Soldier
Nanotechnologies under Contract No. W911NF-07-D0004, and the Unity
through Knowledge Fund Grant Agreement No. 93/11.


\begin{thebibliography}{28}%
\makeatletter
\providecommand \@ifxundefined [1]{%
 \@ifx{#1\undefined}
}%
\providecommand \@ifnum [1]{%
 \ifnum #1\expandafter \@firstoftwo
 \else \expandafter \@secondoftwo
 \fi
}%
\providecommand \@ifx [1]{%
 \ifx #1\expandafter \@firstoftwo
 \else \expandafter \@secondoftwo
 \fi
}%
\providecommand \natexlab [1]{#1}%
\providecommand \enquote  [1]{``#1''}%
\providecommand \bibnamefont  [1]{#1}%
\providecommand \bibfnamefont [1]{#1}%
\providecommand \citenamefont [1]{#1}%
\providecommand \href@noop [0]{\@secondoftwo}%
\providecommand \href [0]{\begingroup \@sanitize@url \@href}%
\providecommand \@href[1]{\@@startlink{#1}\@@href}%
\providecommand \@@href[1]{\endgroup#1\@@endlink}%
\providecommand \@sanitize@url [0]{\catcode `\\12\catcode `\$12\catcode
  `\&12\catcode `\#12\catcode `\^12\catcode `\_12\catcode `\%12\relax}%
\providecommand \@@startlink[1]{}%
\providecommand \@@endlink[0]{}%
\providecommand \url  [0]{\begingroup\@sanitize@url \@url }%
\providecommand \@url [1]{\endgroup\@href {#1}{\urlprefix }}%
\providecommand \urlprefix  [0]{URL }%
\providecommand \Eprint [0]{\href }%
\providecommand \doibase [0]{http://dx.doi.org/}%
\providecommand \selectlanguage [0]{\@gobble}%
\providecommand \bibinfo  [0]{\@secondoftwo}%
\providecommand \bibfield  [0]{\@secondoftwo}%
\providecommand \translation [1]{[#1]}%
\providecommand \BibitemOpen [0]{}%
\providecommand \bibitemStop [0]{}%
\providecommand \bibitemNoStop [0]{.\EOS\space}%
\providecommand \EOS [0]{\spacefactor3000\relax}%
\providecommand \BibitemShut  [1]{\csname bibitem#1\endcsname}%
\let\auto@bib@innerbib\@empty
%</preamble>
\bibitem [{\citenamefont {Landau}\ and\ \citenamefont
  {Lifshitz}(1980)}]{Landau1980}%
  \BibitemOpen
  \bibfield  {author} {\bibinfo {author} {\bibfnamefont {L.~D.}\ \bibnamefont
  {Landau}}\ and\ \bibinfo {author} {\bibfnamefont {E.~M.}\ \bibnamefont
  {Lifshitz}},\ }\href@noop {} {\emph {\bibinfo {title} {{S}tatistical
  {P}hysics, {P}art 2}}}\ (\bibinfo  {publisher} {Pergamon Press},\ \bibinfo
  {year} {1980})\BibitemShut {NoStop}%
\bibitem [{\citenamefont {Hargreaves}(1969)}]{Hargreaves1969}%
  \BibitemOpen
  \bibfield  {author} {\bibinfo {author} {\bibfnamefont {C.}~\bibnamefont
  {Hargreaves}},\ }\href {\doibase 10.1016/0375-9601(69)90264-3} {\bibfield
  {journal} {\bibinfo  {journal} {Phys. Lett. A}\ }\textbf {\bibinfo
  {volume} {30}},\ \bibinfo {pages} {491 } (\bibinfo {year}
  {1969})}\BibitemShut {NoStop}%
\bibitem [{\citenamefont {Narayanaswamy}\ \emph {et~al.}(2008)\citenamefont
  {Narayanaswamy}, \citenamefont {Shen},\ and\ \citenamefont
  {Chen}}]{Narayanaswamy2008a}%
  \BibitemOpen
  \bibfield  {author} {\bibinfo {author} {\bibfnamefont {A.}~\bibnamefont
  {Narayanaswamy}}, \bibinfo {author} {\bibfnamefont {S.}~\bibnamefont {Shen}},
  \ and\ \bibinfo {author} {\bibfnamefont {G.}~\bibnamefont {Chen}},\ }\href
  {\doibase 10.1103/PhysRevB.78.115303} {\bibfield  {journal} {\bibinfo
  {journal} {Phys. Rev. B}\ }\textbf {\bibinfo {volume} {78}},\ \bibinfo
  {pages} {115303} (\bibinfo {year} {2008})}\BibitemShut {NoStop}%
\bibitem [{\citenamefont {Shen}\ \emph {et~al.}(2009)\citenamefont {Shen},
  \citenamefont {Narayanaswamy},\ and\ \citenamefont {Chen}}]{Shen2009a}%
  \BibitemOpen
  \bibfield  {author} {\bibinfo {author} {\bibfnamefont {S.}~\bibnamefont
  {Shen}}, \bibinfo {author} {\bibfnamefont {A.}~\bibnamefont {Narayanaswamy}},
  \ and\ \bibinfo {author} {\bibfnamefont {G.}~\bibnamefont {Chen}},\ }\href
  {\doibase 10.1021/nl901208v} {\bibfield  {journal} {\bibinfo  {journal} {Nano
  Lett.}\ }\textbf {\bibinfo {volume} {9}},\ \bibinfo {pages} {2909}
  (\bibinfo {year} {2009})}\BibitemShut {NoStop}%
\bibitem [{\citenamefont {Rousseau}\ \emph {et~al.}(2009)\citenamefont
  {Rousseau}, \citenamefont {Siria}, \citenamefont {Jourdan}, \citenamefont
  {Volz}, \citenamefont {Comin}, \citenamefont {Chevrier},\ and\ \citenamefont
  {Greffet}}]{Rousseau2009}%
  \BibitemOpen
  \bibfield  {author} {\bibinfo {author} {\bibfnamefont {E.}~\bibnamefont
  {Rousseau}}, \bibinfo {author} {\bibfnamefont {A.}~\bibnamefont {Siria}},
  \bibinfo {author} {\bibfnamefont {G.}~\bibnamefont {Jourdan}}, \bibinfo
  {author} {\bibfnamefont {S.}~\bibnamefont {Volz}}, \bibinfo {author}
  {\bibfnamefont {F.}~\bibnamefont {Comin}}, \bibinfo {author} {\bibfnamefont
  {J.}~\bibnamefont {Chevrier}}, \ and\ \bibinfo {author} {\bibfnamefont
  {J.-J.}\ \bibnamefont {Greffet}},\ }\href {\doibase 10.1038/nphoton.2009.144}
  {\bibfield  {journal} {\bibinfo  {journal} {Nat. Photonics}\ }\textbf
  {\bibinfo {volume} {3}},\ \bibinfo {pages} {514} (\bibinfo {year}
  {2009})}\BibitemShut {NoStop}%
\bibitem [{\citenamefont {Joulain}\ \emph {et~al.}(2005)\citenamefont
  {Joulain}, \citenamefont {Mulet}, \citenamefont {Marquier}, \citenamefont
  {Carminati},\ and\ \citenamefont {Greffet}}]{Joulain2005}%
  \BibitemOpen
  \bibfield  {author} {\bibinfo {author} {\bibfnamefont {K.}~\bibnamefont
  {Joulain}}, \bibinfo {author} {\bibfnamefont {J.-P.}\ \bibnamefont {Mulet}},
  \bibinfo {author} {\bibfnamefont {F.}~\bibnamefont {Marquier}}, \bibinfo
  {author} {\bibfnamefont {R.}~\bibnamefont {Carminati}}, \ and\ \bibinfo
  {author} {\bibfnamefont {J.-J.}\ \bibnamefont {Greffet}},\ }\href {\doibase
  10.1016/j.surfrep.2004.12.002} {\bibfield  {journal} {\bibinfo  {journal}
  {Surf. Sci. Rep.}\ }\textbf {\bibinfo {volume} {57}},\ \bibinfo
  {pages} {59} (\bibinfo {year} {2005})}\BibitemShut {NoStop}%
\bibitem [{\citenamefont {Novoselov}\ \emph {et~al.}(2004)\citenamefont
  {Novoselov}, \citenamefont {Geim}, \citenamefont {Morozov}, \citenamefont
  {Jiang}, \citenamefont {Zhang}, \citenamefont {Dubonos}, \citenamefont
  {Grigorieva},\ and\ \citenamefont {Firsov}}]{Novoselov2004}%
  \BibitemOpen
  \bibfield  {author} {\bibinfo {author} {\bibfnamefont {K.~S.}\ \bibnamefont
  {Novoselov}}, \bibinfo {author} {\bibfnamefont {A.~K.}\ \bibnamefont {Geim}},
  \bibinfo {author} {\bibfnamefont {S.~V.}\ \bibnamefont {Morozov}}, \bibinfo
  {author} {\bibfnamefont {D.}~\bibnamefont {Jiang}}, \bibinfo {author}
  {\bibfnamefont {Y.}~\bibnamefont {Zhang}}, \bibinfo {author} {\bibfnamefont
  {S.~V.}\ \bibnamefont {Dubonos}}, \bibinfo {author} {\bibfnamefont {I.~V.}\
  \bibnamefont {Grigorieva}}, \ and\ \bibinfo {author} {\bibfnamefont {A.~A.}\
  \bibnamefont {Firsov}},\ }\href {\doibase 10.1126/science.1102896} {\bibfield
   {journal} {\bibinfo  {journal} {Science}\ }\textbf {\bibinfo {volume}
  {306}},\ \bibinfo {pages} {666} (\bibinfo {year} {2004})}\BibitemShut
  {NoStop}%
\bibitem [{\citenamefont {Jablan}\ \emph {et~al.}(2009)\citenamefont {Jablan},
  \citenamefont {Buljan},\ and\ \citenamefont {Solja\ifmmode \check{c}\else
  \v{c}\fi{}i\ifmmode~\acute{c}\else \'{c}\fi{}}}]{Jablan2009}%
  \BibitemOpen
  \bibfield  {author} {\bibinfo {author} {\bibfnamefont {M.}~\bibnamefont
  {Jablan}}, \bibinfo {author} {\bibfnamefont {H.}~\bibnamefont {Buljan}}, \
  and\ \bibinfo {author} {\bibfnamefont {M.}~\bibnamefont {Solja\ifmmode
  \check{c}\else \v{c}\fi{}i\ifmmode~\acute{c}\else \'{c}\fi{}}},\ }\href
  {\doibase 10.1103/PhysRevB.80.245435} {\bibfield  {journal} {\bibinfo
  {journal} {Phys. Rev. B}\ }\textbf {\bibinfo {volume} {80}},\ \bibinfo
  {pages} {245435} (\bibinfo {year} {2009})}\BibitemShut {NoStop}%
\bibitem [{\citenamefont {Wunsch}\ \emph {et~al.}(2006)\citenamefont {Wunsch},
  \citenamefont {Stauber}, \citenamefont {Sols},\ and\ \citenamefont
  {Guinea}}]{Wunsch2006}%
  \BibitemOpen
  \bibfield  {author} {\bibinfo {author} {\bibfnamefont {B.}~\bibnamefont
  {Wunsch}}, \bibinfo {author} {\bibfnamefont {T.}~\bibnamefont {Stauber}},
  \bibinfo {author} {\bibfnamefont {F.}~\bibnamefont {Sols}}, \ and\ \bibinfo
  {author} {\bibfnamefont {F.}~\bibnamefont {Guinea}},\ }\href
  {http://stacks.iop.org/1367-2630/8/i=12/a=318} {\bibfield  {journal}
  {\bibinfo  {journal} {New J. Phys.}\ }\textbf {\bibinfo {volume}
  {8}},\ \bibinfo {pages} {318} (\bibinfo {year} {2006})}\BibitemShut {NoStop}%
\bibitem [{\citenamefont {Hwang}\ and\ \citenamefont
  {Das~Sarma}(2007)}]{Hwang2007}%
  \BibitemOpen
  \bibfield  {author} {\bibinfo {author} {\bibfnamefont {E.~H.}\ \bibnamefont
  {Hwang}}\ and\ \bibinfo {author} {\bibfnamefont {S.}~\bibnamefont
  {Das~Sarma}},\ }\href {\doibase 10.1103/PhysRevB.75.205418} {\bibfield
  {journal} {\bibinfo  {journal} {Phys. Rev. B}\ }\textbf {\bibinfo {volume}
  {75}},\ \bibinfo {pages} {205418} (\bibinfo {year} {2007})}\BibitemShut
  {NoStop}%
\bibitem [{\citenamefont {Ju}\ \emph {et~al.}(2011)\citenamefont {Ju},
  \citenamefont {Geng}, \citenamefont {Horng}, \citenamefont {Girit},
  \citenamefont {Martin}, \citenamefont {Hao}, \citenamefont {Bechtel},
  \citenamefont {Liang}, \citenamefont {Zettl}, \citenamefont {Shen},\ and\
  \citenamefont {Wang}}]{Ju2011}%
  \BibitemOpen
  \bibfield  {author} {\bibinfo {author} {\bibfnamefont {L.}~\bibnamefont
  {Ju}}, \bibinfo {author} {\bibfnamefont {B.}~\bibnamefont {Geng}}, \bibinfo
  {author} {\bibfnamefont {J.}~\bibnamefont {Horng}}, \bibinfo {author}
  {\bibfnamefont {C.}~\bibnamefont {Girit}}, \bibinfo {author} {\bibfnamefont
  {M.}~\bibnamefont {Martin}}, \bibinfo {author} {\bibfnamefont
  {Z.}~\bibnamefont {Hao}}, \bibinfo {author} {\bibfnamefont {H.~A.}\
  \bibnamefont {Bechtel}}, \bibinfo {author} {\bibfnamefont {X.}~\bibnamefont
  {Liang}}, \bibinfo {author} {\bibfnamefont {A.}~\bibnamefont {Zettl}},
  \bibinfo {author} {\bibfnamefont {Y.~R.}\ \bibnamefont {Shen}}, \ and\
  \bibinfo {author} {\bibfnamefont {F.}~\bibnamefont {Wang}},\ }\href {\doibase
  10.1038/nnano.2011.146} {\bibfield  {journal} {\bibinfo  {journal} {Nat
  Nano}\ }\textbf {\bibinfo {volume} {6}},\ \bibinfo {pages} {630} (\bibinfo
  {year} {2011})}\BibitemShut {NoStop}%
\bibitem [{\citenamefont {Vakil}\ and\ \citenamefont
  {Engheta}(2011)}]{Vakil2011}%
  \BibitemOpen
  \bibfield  {author} {\bibinfo {author} {\bibfnamefont {A.}~\bibnamefont
  {Vakil}}\ and\ \bibinfo {author} {\bibfnamefont {N.}~\bibnamefont
  {Engheta}},\ }\href {\doibase 10.1126/science.1202691} {\bibfield  {journal}
  {\bibinfo  {journal} {Science}\ }\textbf {\bibinfo {volume} {332}},\ \bibinfo
  {pages} {1291} (\bibinfo {year} {2011})}\BibitemShut {NoStop}%
\bibitem [{\citenamefont {Rana}(2011)}]{Rana2011}%
  \BibitemOpen
  \bibfield  {author} {\bibinfo {author} {\bibfnamefont {F.}~\bibnamefont
  {Rana}},\ }\href {\doibase 10.1038/nnano.2011.170} {\bibfield  {journal}
  {\bibinfo  {journal} {Nat. Nano}\ }\textbf {\bibinfo {volume}
  {6}},\ \bibinfo {pages} {611} (\bibinfo {year} {2011})}\BibitemShut {NoStop}%
\bibitem [{\citenamefont {Persson}\ and\ \citenamefont
  {Ueba}(2010)}]{Persson2010}%
  \BibitemOpen
  \bibfield  {author} {\bibinfo {author} {\bibfnamefont {B.~N.~J.}\
  \bibnamefont {Persson}}\ and\ \bibinfo {author} {\bibfnamefont
  {H.}~\bibnamefont {Ueba}},\ }\href
  {http://stacks.iop.org/0953-8984/22/i=46/a=462201} {\bibfield  {journal}
  {\bibinfo  {journal} {J. Phys. Condens. Matter}\ }\textbf
  {\bibinfo {volume} {22}},\ \bibinfo {pages} {462201} (\bibinfo {year}
  {2010})}\BibitemShut {NoStop}%
\bibitem [{\citenamefont {Volokitin}\ and\ \citenamefont
  {Persson}(2011)}]{Volokitin2011}%
  \BibitemOpen
  \bibfield  {author} {\bibinfo {author} {\bibfnamefont {A.~I.}\ \bibnamefont
  {Volokitin}}\ and\ \bibinfo {author} {\bibfnamefont {B.~N.~J.}\ \bibnamefont
  {Persson}},\ }\href {\doibase 10.1103/PhysRevB.83.241407} {\bibfield
  {journal} {\bibinfo  {journal} {Phys. Rev. B}\ }\textbf {\bibinfo {volume}
  {83}},\ \bibinfo {pages} {241407} (\bibinfo {year} {2011})}\BibitemShut
  {NoStop}%
\bibitem [{\citenamefont {Ilic}\ \emph {et~al.}(2012)\citenamefont {Ilic},
  \citenamefont {Jablan}, \citenamefont {Joannopoulos}, \citenamefont
  {Celanovic},\ and\ \citenamefont {Solja\v{c}i\'{c}}}]{Ilic2012}%
  \BibitemOpen
  \bibfield  {author} {\bibinfo {author} {\bibfnamefont {O.}~\bibnamefont
  {Ilic}}, \bibinfo {author} {\bibfnamefont {M.}~\bibnamefont {Jablan}},
  \bibinfo {author} {\bibfnamefont {J.~D.}\ \bibnamefont {Joannopoulos}},
  \bibinfo {author} {\bibfnamefont {I.}~\bibnamefont {Celanovic}}, \ and\
  \bibinfo {author} {\bibfnamefont {M.}~\bibnamefont {Solja\v{c}i\'{c}}},\
  }\href {\doibase 10.1364/OE.20.00A366} {\bibfield  {journal} {\bibinfo
  {journal} {Opt. Express}\ }\textbf {\bibinfo {volume} {20}},\ \bibinfo
  {pages} {A366} (\bibinfo {year} {2012})}\BibitemShut {NoStop}%
\bibitem [{\citenamefont {Rytov}\ \emph {et~al.}(1987)\citenamefont {Rytov},
  \citenamefont {Kratsov},\ and\ \citenamefont {Tatarskii}}]{Rytov1987}%
  \BibitemOpen
  \bibfield  {author} {\bibinfo {author} {\bibfnamefont {S.}~\bibnamefont
  {Rytov}}, \bibinfo {author} {\bibfnamefont {Y.~A.}\ \bibnamefont {Kratsov}},
  \ and\ \bibinfo {author} {\bibfnamefont {V.~I.}\ \bibnamefont {Tatarskii}},\
  }\href@noop {} {\emph {\bibinfo {title} {{P}rinciples of {S}tatistical
  {R}adiophysics}}}\ (\bibinfo  {publisher} {Springer-Verlag},\ \bibinfo {year}
  {1987})\BibitemShut {NoStop}%
\bibitem [{\citenamefont {Haus}(1961)}]{Haus1961}%
  \BibitemOpen
  \bibfield  {author} {\bibinfo {author} {\bibfnamefont {H.~A.}\ \bibnamefont
  {Haus}},\ }\href {\doibase DOI:10.1063/1.1736031} {\bibfield  {journal}
  {\bibinfo  {journal} {J. Appl. Phys.}\ }\textbf {\bibinfo
  {volume} {32}},\ \bibinfo {pages} {493} (\bibinfo {year} {1961})}\BibitemShut
  {NoStop}%
\bibitem [{\citenamefont {Sipe}(1987)}]{Sipe1987}%
  \BibitemOpen
  \bibfield  {author} {\bibinfo {author} {\bibfnamefont {J.~E.}\ \bibnamefont
  {Sipe}},\ }\href {\doibase 10.1364/JOSAB.4.000481} {\bibfield  {journal}
  {\bibinfo  {journal} {J. Opt. Soc. Am. B}\ }\textbf {\bibinfo {volume} {4}},\
  \bibinfo {pages} {481} (\bibinfo {year} {1987})}\BibitemShut {NoStop}%
\bibitem [{\citenamefont {Pendry}(1999)}]{Pendry1999a}%
  \BibitemOpen
  \bibfield  {author} {\bibinfo {author} {\bibfnamefont {J.}~\bibnamefont
  {Pendry}},\ }\href@noop {} {\bibfield  {journal} {\bibinfo  {journal}
  {J. Phys. Condens. Matter}\ }\textbf {\bibinfo {volume} {11}},\
  \bibinfo {pages} {6621} (\bibinfo {year} {1999})}\BibitemShut {NoStop}%
\bibitem [{\citenamefont {Ben-Abdallah}\ \emph {et~al.}(2009)\citenamefont
  {Ben-Abdallah}, \citenamefont {Joulain}, \citenamefont {Drevillon},\ and\
  \citenamefont {Domingues}}]{Ben-Abdallah2009}%
  \BibitemOpen
  \bibfield  {author} {\bibinfo {author} {\bibfnamefont {P.}~\bibnamefont
  {Ben-Abdallah}}, \bibinfo {author} {\bibfnamefont {K.}~\bibnamefont
  {Joulain}}, \bibinfo {author} {\bibfnamefont {J.}~\bibnamefont {Drevillon}},
  \ and\ \bibinfo {author} {\bibfnamefont {G.}~\bibnamefont {Domingues}},\
  }\href {\doibase 10.1063/1.3204481} {\bibfield  {journal} {\bibinfo
  {journal} {J. Appl. Phys.}\ }\textbf {\bibinfo {volume} {106}},\
  \bibinfo {eid} {044306} (\bibinfo {year} {2009})}\BibitemShut {NoStop}%
\bibitem [{\citenamefont {Mulet}\ \emph {et~al.}(2002)\citenamefont {Mulet},
  \citenamefont {Joulain}, \citenamefont {Carminati},\ and\ \citenamefont
  {Greffet}}]{Mulet2002}%
  \BibitemOpen
  \bibfield  {author} {\bibinfo {author} {\bibfnamefont {J.-P.}\ \bibnamefont
  {Mulet}}, \bibinfo {author} {\bibfnamefont {K.}~\bibnamefont {Joulain}},
  \bibinfo {author} {\bibfnamefont {R.}~\bibnamefont {Carminati}}, \ and\
  \bibinfo {author} {\bibfnamefont {J.-J.}\ \bibnamefont {Greffet}},\
  }\href@noop {} {\bibfield  {journal} {\bibinfo  {journal} {Microscale
  Thermophys. Eng.}\ }\textbf {\bibinfo {volume} {6}},\ \bibinfo
  {pages} {209} (\bibinfo {year} {2002})}\BibitemShut {NoStop}%
\bibitem [{\citenamefont {Narayanaswamy}\ and\ \citenamefont
  {Chen}(2008)}]{Narayanaswamy2008}%
  \BibitemOpen
  \bibfield  {author} {\bibinfo {author} {\bibfnamefont {A.}~\bibnamefont
  {Narayanaswamy}}\ and\ \bibinfo {author} {\bibfnamefont {G.}~\bibnamefont
  {Chen}},\ }\href {\doibase 10.1103/PhysRevB.77.075125} {\bibfield  {journal}
  {\bibinfo  {journal} {Phys. Rev. B}\ }\textbf {\bibinfo {volume} {77}},\
  \bibinfo {pages} {075125} (\bibinfo {year} {2008})}\BibitemShut {NoStop}%
\bibitem [{\citenamefont {Rodriguez}\ \emph {et~al.}(2011)\citenamefont
  {Rodriguez}, \citenamefont {Ilic}, \citenamefont {Bermel}, \citenamefont
  {Celanovic}, \citenamefont {Joannopoulos}, \citenamefont {Solja\ifmmode
  \check{c}\else \v{c}\fi{}i\ifmmode~\acute{c}\else \'{c}\fi{}},\ and\
  \citenamefont {Johnson}}]{Rodriguez2011}%
  \BibitemOpen
  \bibfield  {author} {\bibinfo {author} {\bibfnamefont {A.~W.}\ \bibnamefont
  {Rodriguez}}, \bibinfo {author} {\bibfnamefont {O.}~\bibnamefont {Ilic}},
  \bibinfo {author} {\bibfnamefont {P.}~\bibnamefont {Bermel}}, \bibinfo
  {author} {\bibfnamefont {I.}~\bibnamefont {Celanovic}}, \bibinfo {author}
  {\bibfnamefont {J.~D.}\ \bibnamefont {Joannopoulos}}, \bibinfo {author}
  {\bibfnamefont {M.}~\bibnamefont {Solja\ifmmode \check{c}\else
  \v{c}\fi{}i\ifmmode~\acute{c}\else \'{c}\fi{}}}, \ and\ \bibinfo {author}
  {\bibfnamefont {S.~G.}\ \bibnamefont {Johnson}},\ }\href {\doibase
  10.1103/PhysRevLett.107.114302} {\bibfield  {journal} {\bibinfo  {journal}
  {Phys. Rev. Lett.}\ }\textbf {\bibinfo {volume} {107}},\ \bibinfo {pages}
  {114302} (\bibinfo {year} {2011})}\BibitemShut {NoStop}%
%\bibitem [{der()}]{derivationNote}%
  %\BibitemOpen
  %\href@noop {} {}\bibinfo {note} {It is interesting to note that the
  %expression for the near-field spectral transfer function between two graphene
  %sheets is the same as that for two semi-infinite slabs. For complete
  %derivation for the case of graphene see [16]}\BibitemShut {NoStop}%
\bibitem [{\citenamefont {Falkovsky}(2008)}]{Falkovsky2008}%
  \BibitemOpen
  \bibfield  {author} {\bibinfo {author} {\bibfnamefont {L.~A.}\ \bibnamefont
  {Falkovsky}},\ }\href {http://stacks.iop.org/1742-6596/129/i=1/a=012004}
  {\bibfield  {journal} {\bibinfo  {journal} {J. Phys. Conf. Ser.}\ }\textbf {\bibinfo {volume} {129}},\ \bibinfo {pages} {012004}
  (\bibinfo {year} {2008})}\BibitemShut {NoStop}%
\bibitem [{\citenamefont {Mikhailov}\ and\ \citenamefont
  {Ziegler}(2007)}]{Mikhailov2007}%
  \BibitemOpen
  \bibfield  {author} {\bibinfo {author} {\bibfnamefont {S.~A.}\ \bibnamefont
  {Mikhailov}}\ and\ \bibinfo {author} {\bibfnamefont {K.}~\bibnamefont
  {Ziegler}},\ }\href {\doibase 10.1103/PhysRevLett.99.016803} {\bibfield
  {journal} {\bibinfo  {journal} {Phys. Rev. Lett.}\ }\textbf {\bibinfo
  {volume} {99}},\ \bibinfo {pages} {016803} (\bibinfo {year}
  {2007})}\BibitemShut {NoStop}%
\bibitem [{\citenamefont {Gabor}\ \emph {et~al.}(2011)\citenamefont {Gabor},
  \citenamefont {Song}, \citenamefont {Ma}, \citenamefont {Nair}, \citenamefont
  {Taychatanapat}, \citenamefont {Watanabe}, \citenamefont {Taniguchi},
  \citenamefont {Levitov},\ and\ \citenamefont {Jarillo-Herrero}}]{Gabor2011}%
  \BibitemOpen
  \bibfield  {author} {\bibinfo {author} {\bibfnamefont {N.~M.}\ \bibnamefont
  {Gabor}}, \bibinfo {author} {\bibfnamefont {J.~C.~W.}\ \bibnamefont {Song}},
  \bibinfo {author} {\bibfnamefont {Q.}~\bibnamefont {Ma}}, \bibinfo {author}
  {\bibfnamefont {N.~L.}\ \bibnamefont {Nair}}, \bibinfo {author}
  {\bibfnamefont {T.}~\bibnamefont {Taychatanapat}}, \bibinfo {author}
  {\bibfnamefont {K.}~\bibnamefont {Watanabe}}, \bibinfo {author}
  {\bibfnamefont {T.}~\bibnamefont {Taniguchi}}, \bibinfo {author}
  {\bibfnamefont {L.~S.}\ \bibnamefont {Levitov}}, \ and\ \bibinfo {author}
  {\bibfnamefont {P.}~\bibnamefont {Jarillo-Herrero}},\ }\href {\doibase
  10.1126/science.1211384} {\bibfield  {journal} {\bibinfo  {journal}
  {Science}\ }\textbf {\bibinfo {volume} {334}},\ \bibinfo {pages} {648}
  (\bibinfo {year} {2011})}\BibitemShut {NoStop}%
\end{thebibliography}
\end{document}